\documentclass[journal=jacsat,manuscript=article]{achemso}

\usepackage[version=3]{mhchem} 
\usepackage{float}

\author{Oren Elishav}
\altaffiliation{O.E. and R.P. contributed equally to this work}
\author{Roy Podgaetsky}
\altaffiliation{O.E. and R.P. contributed equally to this work}
\affiliation
{School of Chemistry, Tel Aviv University, Tel Aviv 6997801, Israel}
\author{Olga Meikler}
\affiliation
{Rafael Ltd., P.O. Box 2250, Haifa 3102102, Israel}
\author{Barak Hirshberg}
\email{hirshb@tauex.tau.ac.il}
\affiliation
{School of Chemistry, Tel Aviv University, Tel Aviv 6997801, Israel}
\alsoaffiliation[Sackler]
{The Center for Computational Molecular and Materials Science, Tel Aviv University, Tel Aviv 6997801, Israel.}
\alsoaffiliation[Ratner]
{The Ratner Center for Single Molecule Science, Tel Aviv University, Tel Aviv 6997801, Israel.}

\title
  {Collective Variables for Conformational Polymorphism in Molecular Crystals}

\abbreviations{}
\keywords{}

\begin{document}


\begin{abstract}
Controlling polymorphism in molecular crystals is crucial in the pharmaceutical, dye, and pesticide industries. However, its theoretical description is extremely challenging, due to the associated long timescales ($ > 1 \, \mu s$). We present an efficient procedure for identifying collective variables that promote transitions between conformational polymorphs in molecular dynamics simulations. It involves applying a simple dimensionality reduction algorithm to data from short ($\sim ps$) simulations of the isolated conformers that correspond to each polymorph. We demonstrate the utility of our method in the challenging case of the important energetic material, CL-20, which has three anhydrous conformational polymorphs at ambient pressure. Using these collective variables in Metadynamics simulations, we observe transitions between all solid polymorphs in the biased trajectories. We reconstruct the free energy surface and identify previously unknown defect and intermediate forms in the transition from one known polymorph to another. Our method provides insights into complex conformational polymorphic transitions of flexible molecular crystals.
\end{abstract}

Polymorphism in molecular solids is of great importance in the development of new pesticides, pigments, pharmaceuticals and energetic materials.\cite{Nogueira2020,Liu2018}
Simulations can help design  procedures to isolate a desired crystalline form and understand the underlying transition mechanisms.\cite{Schneider2016,Hamad2006} However, molecular dynamics (MD) simulations of phase transitions in solids are challenging, since they occur on a time scale longer than $ 1 \,\mu s$, which is unreachable using standard methods.\cite{Bussi2020} 
Enhanced sampling algorithms, such as Metadynamics (MetaD), bias the simulation to amplify the occurrence of such rare events in the trajectories. Many of them rely on identifying collective variables (CVs) that, ideally, accelerate the sampling of the slowest modes involved in the process.\cite{Invernizzi2019} 

Finding good CVs for polymorphism is challenging but progress has been made for stiff molecules.\cite{Raiteri2005,Schneider2016,Pipolo2017,Piaggi2018,Gimondi2017} For example, \citeauthor{Piaggi2018} recently constructed a CV based on distance and relative orientation between neighboring molecules to enhance the polymorphic transitions of urea and naphthalene.\cite{Piaggi2018} 
\citeauthor{Gimondi2017} used a CV that reflects the local environment around CO${_2}$ molecules in the solid to promote its polymorphism.\cite{Gimondi2017} 
These simulations uncovered new polymorphs, revealed defect phases, and provided their interconversion barriers. 
In conformational polymorphism, the crystalline forms differ by the conformation of the constituents molecules and not just by their relative orientation or lattice parameters. This can significantly affect the chemical and physical properties of the solid.\cite{Cruz-Cabeza2014} The added complexity of conformational polymorphism poses a challenge in finding optimal CVs.\cite{Bonati2020} Therefore, simulations of polymorphic transitions involving conformation changes are much rarer. 

Recent studies showed the benefits of using data-driven approaches to identify suitable CVs.\cite{Bonati2020,Keith2021,Bonati2019,Bonati2021,Sidky2020,Zhang2019,Gasparotto2020,Rizzi2019} 
For example, \citeauthor{Mendels2018} used harmonic linear discriminant analysis (HLDA) to obtain CVs that describe the phase transition from the liquid to a superionic phase of AgI.\cite{Mendels2018} \citeauthor{Piccini2018} extended the HLDA method to address multiple metastable states (multi-class HLDA, MC-HLDA) and applied it to obtain the free energy surface (FES) of chemical reactions.\cite{Piccini2018} 
Here, we propose a procedure to obtain CVs that are able to enhance conformational polymorphic transitions in simulations of molecular crystals. It is based on applying MC-HLDA to data obtained solely from short MD simulations of the isolated conformers that correspond to each polymorph. It is also the first example of applying HLDA or its extensions to molecular crystals, to the best of our knowledge.


As a concrete example, we focus on an important energetic material, Hexanitrohexaazaisowurtzitane (CL-20), because it is a very challenging system (four molecules in the unit cell, 36 atoms per molecule) that demonstrates a rich conformational polymorphism with five different crystalline forms. The polymorphs \(\beta\)-, \(\gamma\)-, \(\epsilon\)-CL-20, and a hydrate \(\alpha\) form, can be obtained at ambient pressure, while the \(\zeta\) polymorph is stable only at high pressure.\cite{Liu2018} Polymorphism in CL-20 plays a crucial role in its synthesis, storage and aging \cite{Bu2020}. The molecular conformations in the polymorphs differ mainly in the improper angles between the nitro groups and the center cage carbon atoms (see Figure 1). Previous computational studies focused on gas- and solid-phase static calculations\cite{Wang2021,Bao2021,Xu2007,Kholod2007} or the interaction energy of CL-20 with various materials.\cite{Wei2016,Zeng2017,Zhang2018,Hao2020} No MD simulations of polymorphic phase transitions between the ambient CL-20 forms have been performed previously. Below, we first present the procedure for identifying CVs to describe conformational polymorphism in molecular crystals. Then, we apply it to accelerate transitions between the polymorphs of CL-20. Finally, we obtain the FES and identify previously unknown defect and intermediate states in the transition from one metastable form to another.

\begin{figure} [H]
  \includegraphics [scale=0.07] {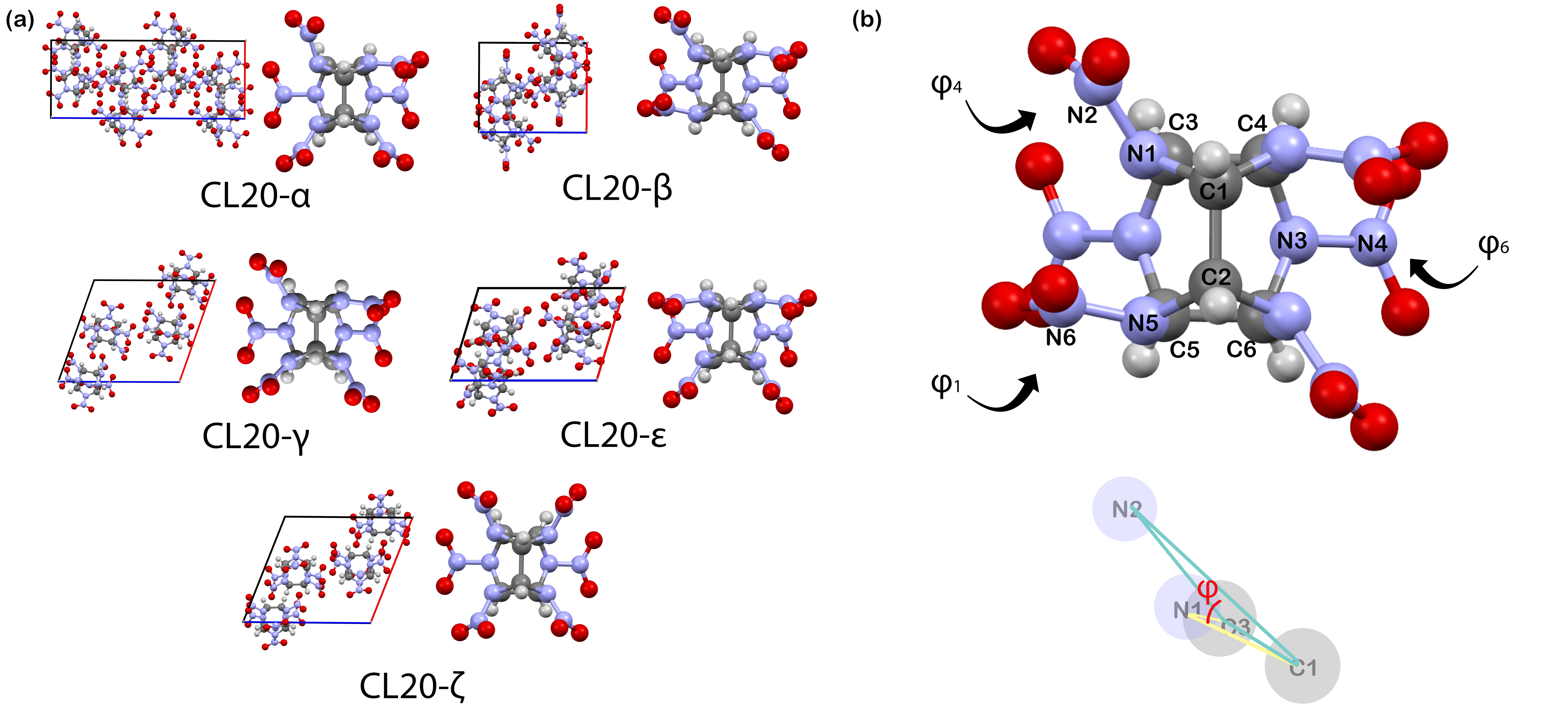}
  \caption{(a) Conformational polymorphs of CL-20, and (b) definition of important improper angles in CL-20, denoted as \(\varphi\)$_1$= N5-C2-C5-N6, \(\varphi\)$_4$= N1-C1-C3-N2, and \(\varphi\)$_6$= N3-C6-C4-N4.}
  \label{fgr:1}
\end{figure}

In the case of conformational polymorphism, we hypothesize that the molecular degrees of freedom would dominate when identifying a suitable collective variable. As a result, we propose the following procedure to obtain the FES for transitions between conformational polymorphs in molecular crystals: 1) Perform short simulations of isolated conformers that correspond to each polymorph. Ensure that no transitions to other conformers occured in the simulations. 2) Identify a CV by performing HLDA on a set of labeled molecular descriptors using data from the simulations above. If there is more than two polymorphs, use MC-HLDA instead. 3) Use the resulting CV in well-tempered Metadynamics (WT-MetaD) simulations starting from the most stable polymorph. If the molecular CV is efficient in promoting transitions between all polymorphs, converge the FES and obtain insight into their relative stability and their interconversion pathways. We test our hypothesis and procedure below for the challenging case of the three ambient, anhydrous polymorphs of CL-20. 

We first outline the computational details used in the specific application of the procedure to CL-20. 
As explained above, the improper angles between the nitro group and the central molecular cage differentiate the three conformational polymorphs (see Figure~\ref{fgr:1}). Therefore, we chose them as descriptors in the MC-HLDA below.
When conducting simulations of the isolated CL-20 conformers at ambient temperature (300K) following step (1) above, we initially observed rapid conformational transitions. This is because, unlike in the solid state, conformation changes are not a rare event in the gas phase. Hence, we performed the short simulations ($\sim70 ps$) of the isolated conformers at a lower temperature (50K). This assured that no conformational transitions occured during step (1), and we only sample fluctuations of a single conformer basin in each trajectory. Next, we used the labeled data as input to MC-HLDA (three classes, one for each polymorph) to obtain two linear combinations of the improper angles as the CV. Then, we set up solid-state simulations starting from the experimental unit cell of the \(\epsilon\) polymorph with periodic boundary conditions at ambient temperature (300K) and pressure (1 atm). Before performing the WT-MetaD simulations, we let the unit cell relax (minimization using a conjugate gradients algorithm) and thermalize for $1.1 \, ns$. During this stage, we verified that no polymorphic transitions occurred. Finally, we performed WT-MetaD simulations using the CV obtained from MC-HLDA.    

The simulations were performed using LAMMPS (30 Jul 2021)\cite{Thompson2022} and PLUMED 2.7.1\cite{Tribello2014,Bonomi2019} with the force-field (FF) of ref. \cite{Bidault2019}. We tested this FF by performing unbiased MD simulations of the \(\beta\)-, \(\gamma\)-, and \(\epsilon\)-forms in the isothermal-isobaric (NPT) ensemble at ambient conditions, which showed good agreement between predicted values using this FF and experimental data for the density (maximal deviation of 3.5\%) and cell parameters (maximal deviation of 6.8\%) of each polymorph (see Table S1). We used a time-step of $1 \, fs$ and performed the simulations at constant temperature and ambient pressure. The full computational details of the calculations can be found in the supporting information. WT-MetaD simulations were performed with a bias factor of 25, The Gaussian hills were deposited every 100 steps and their initial height was 0.2 $kcal \, mol^{-1}$. The Gaussian widths of CV1 and CV2 were 0.08 and 0.13, respectively.  

\begin{figure} [H]
  \includegraphics [scale=0.6] {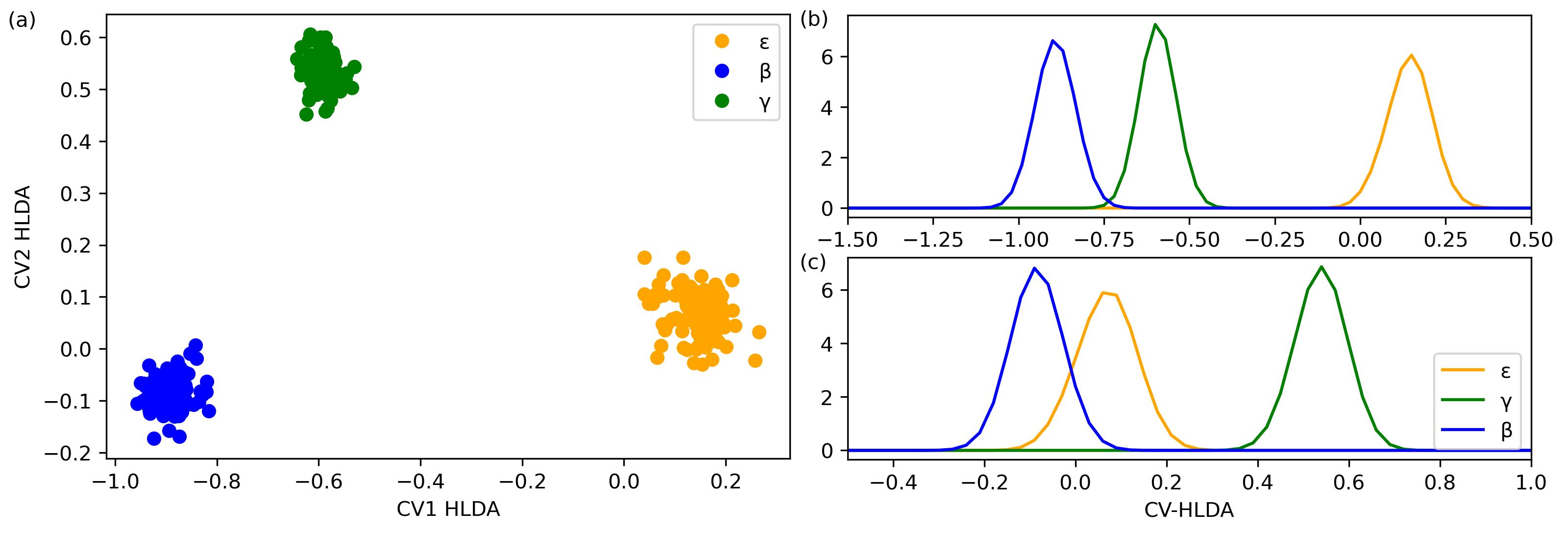} 
  \caption{(a) Scatter plot of data from unbiased simulations of isolated conformers, corresponding to each polymorph of CL-20 polymorphs, in the space of the MC-HLDA CVs. (b) Histogram of HLDA-based CV1 , and (c) HLDA-based CV2 for CL-20 conformers from the same simulations.}
  \label{fgr:2}
\end{figure}

The results of step (1) of the procedure above are given in Figure S1, showing the fluctuations in the six improper angles as a function of time. We find that only three of them, \(\varphi\)$_1$, \(\varphi\)$_4$, and \(\varphi\)$_6$, are substantially different in the three conformers. A preliminary MC-HLDA on all improper angles also showed that the relative weights of \(\varphi\)$_2$, \(\varphi\)$_3$, and \(\varphi\)$_5$ are negligible (Table S2).
As a result, we employed MC-HLDA to generate CVs for conformational polymorphic transitions of CL-20 using only the angles \(\varphi\)$_1$, \(\varphi\)$_4$, and \(\varphi\)$_6$. 
In the MC-HLDA and in Figure S1, the cosines of the improper angles with an offset phase of 1.2 radians were used as descriptors, following Tiwary et al.\cite{Tiwary2016}, to avoid periodicity-related numerical issues.
In the case of a three-class problem, MC-HLDA generates two linear combinations of the descriptors.
A scatter plot of the data from the simulations of the isolated conformers corresponding to the three polymorphs in the two CV space is given in Figure~\ref{fgr:2}(a). We find that the conformer basins are well separated.
The coefficients of each improper angle in the two CVs are given in Table S3.
Their squared value gives the relative contribution for each descriptor in the CV.
We find that the first CV (corresponding to the lowest eigenvalue in the MC-HLDA) has contributions from all three angles while the second CV (corresponding to the second-lowest eigenvalue) is dominated by \(\varphi\)$_1$ and \(\varphi\)$_4$.
The histograms of the fluctuations in both CV during the unbiased simulations of the isolated conformers are given in Panel (b) and (c) of Figure~\ref{fgr:2}.
While at 50K it might look like CV1 is sufficient to separate all three conformers, the fluctuations at 300K for the solid-state simulations (see the next paragraph) are larger and the two CVs are needed to minimize the overlap between the three polymorphs (see Figure S2).


Next, we used WT-MetaD to investigate conformational polymorphic transitions of CL-20 in the solid by enhancing the sampling along the two CVs obtained by MC-HLDA. During a 500 ns biased simulation, many transitions were observed between the three forms (\(\beta\), \(\gamma\), and \(\epsilon\)), as shown in Figures 3(a) and 3(b), plotting the CV values versus time. We confirmed that the transitions in the CV values are accompanied by a transition in the conformation of the CL-20 molecules in the unit cell. 
We also ensured that changes in the CVs are accompanied by transitions in cell parameters (Figure 3c). Remarkably, this confirms our hypothesis that local and molecular CVs are able to drive conformational polymorphic transitions for the complicated case of CL-20. This is done by biasing a molecular conformation transition, which is accompanied by a change in lattice parameters without being baised directly. It is an exciting example where a local CV is able to drive a global phase transition.
In the transitions between the \(\beta\)- and \(\epsilon\)-forms, the obtained cell parameters agree with the experimental values. The transition to the \(\gamma\)-form resulted in slightly different lattice constants than the experimental values (see Table S4). 
We also observed defect and intermediate forms based on the CV values and cell parameters given in Figure 3, whose structure will be analyzed in detail shortly. 
First, we obtain the FES and then analyze the relative stability of all polymorphs, intermediates and defect states.

\begin{figure} [H]
  \includegraphics [scale=0.65] {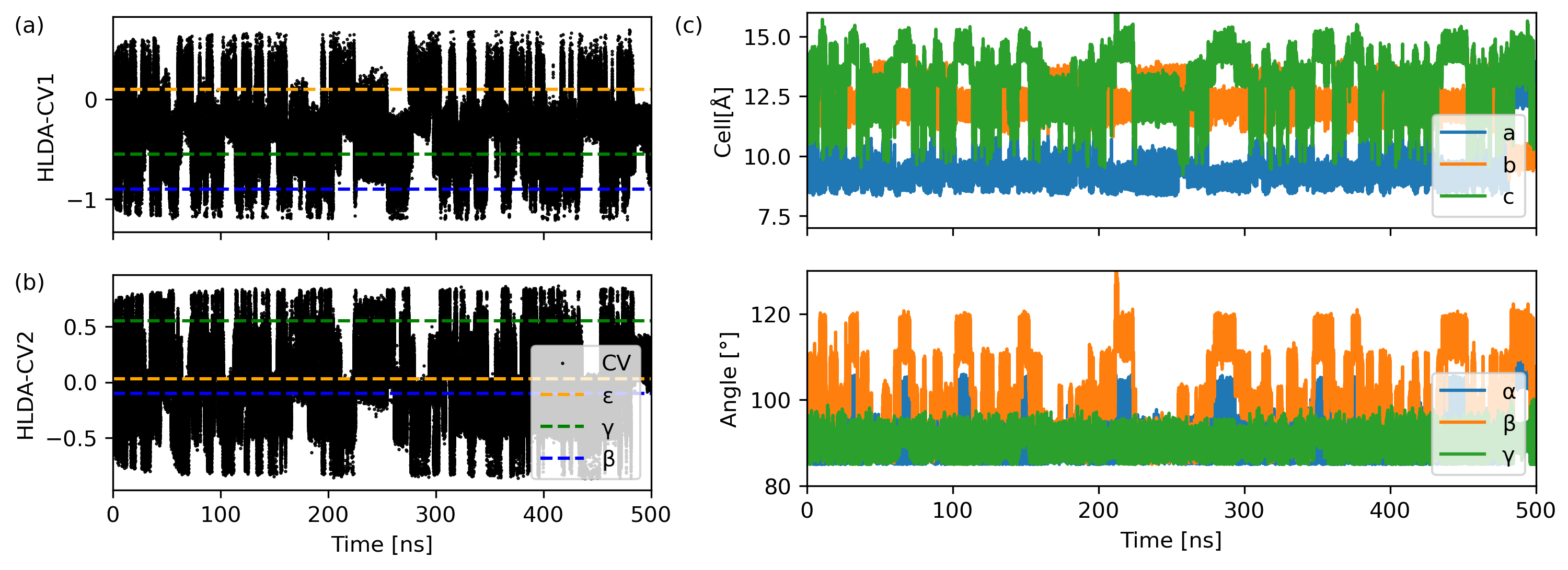}
  \caption{Time evolution during WTMetaD simulation of 500 ns of (a) CV1, (b) CV2, and (c) lattice parameters.
  }
  \label{fgr:3}
\end{figure}

\begin{figure} [H]
  \includegraphics [scale=0.65] {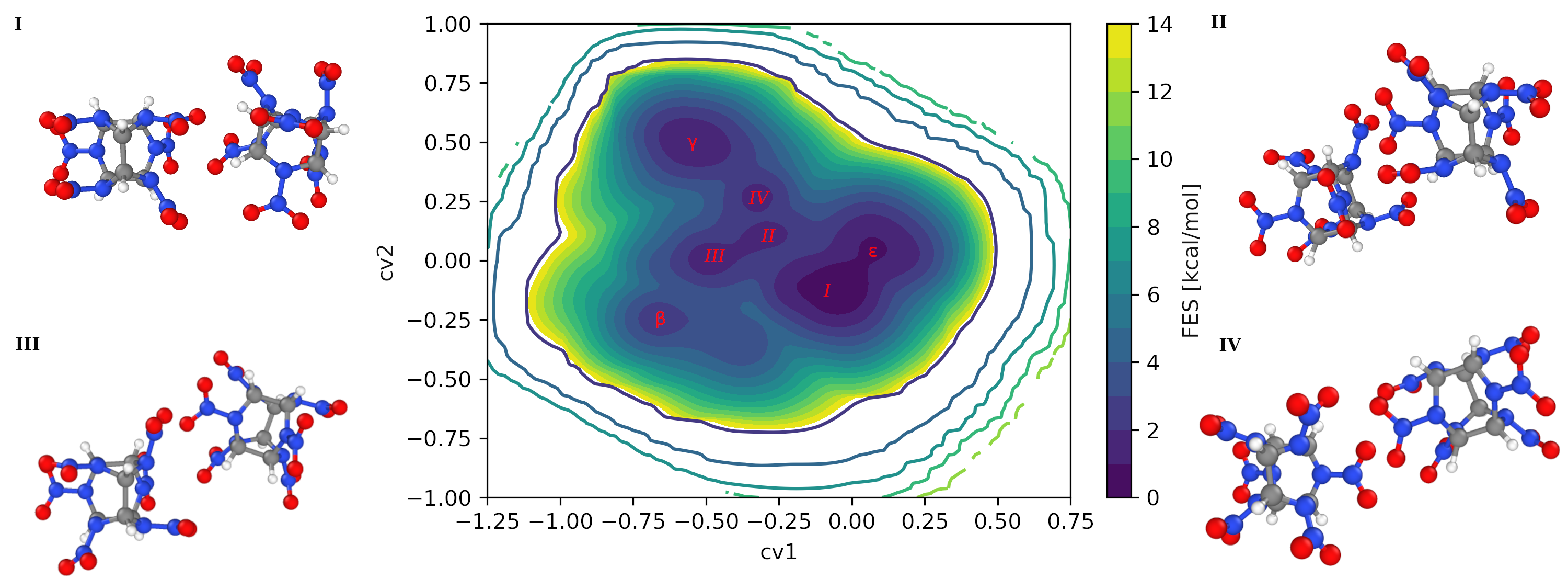}
  \caption{Free energy surface along CV1 and CV2 and molecular conformations of the defected forms.}
  \label{fgr:4}
\end{figure}

The biased simulation resulted in many recrossings between the three polymorphs, thus allowing the reconstruction of the FES using a reweighting procedure.\cite{Bonomi2009,Tiwary2015,Bussi2020} The FES is presented as a function of the two MC-HLDA-based CVs in Figure~\ref{fgr:4}. The locations of the \(\beta\)-, \(\gamma\)-, and \(\epsilon\)-forms' minima in the CVs space are close to the predicted ones from the unbiased isolated conformers simulations (Figure 2). We obtain the correct thermodynamic stability order for the various polymorphs ($ \epsilon>\gamma>\beta $), consistent with previous computational and experimental studies.\cite{Liu2018,Wang2021,Foltz1994} 
We verified that the difference in free energy between \(\epsilon\)- and \(\gamma\)-CL20 is converged (see Figure S3b). The average error in the FES projected onto CV1 is 0.08 $kcal \, mol^{-1}$ (Figure S3a).
As indicated before, during the biased simulation, we also identified several defect and intermediate forms that can be seen as local minima in FES (Figure 4). 
The three defect forms are composed of CL-20 molecules with a mixed molecular conformation of \(\beta\) and \(\gamma\) (II and III) and \(\epsilon\) and \(\gamma\) (IV), and cell parameters as in Table S4. A fourth, intermediate, form has a new hybrid molecular conformation (I). Two of the CL-20 molecules in the unit cell are of \(\epsilon\) structure, and the other two are of a new orientation, not corresponding to any of the previously known polymorphs (see Figure 4).
To the best of our knowledge, it has not been previously reported and it would be an exciting challenge to isolate it experimentally. 
To verify our prediction, we confirmed in unbiased simulations of 1 ns that the intermediate form does not spontaneously transform to one of the stable polymorphs and is indeed a metastable structure. Finally, we preformed four independent simulations and obtained similar FES (see Figure S4). In some cases, the defect forms II-IV are not observed in the FES. However, form I is observed in all the FES, supporting its classification as a metastable intermediate and not as an unstable defect form.

A limitation of our approach is that we had to reduce the temperature to 50 K in the simulations of step (1) to avoid transitions between the conformers in the gas phase. As a result, the input data to MC-HLDA includes smaller fluctuations than those observed in the solid-state simulations at 300 K. Therefore, the separation of the unit cell simulations in the two-dimensional CV space is less profound, showing some overlap between the basins (see Figure S2). Still, it is sufficient to drive polymorphic transitions in the solid, as described above. Due to the small overlap between the basins in the resulting FES, the prediction of the relative stability of the polymorphs is more reliable than their inter-conversion barriers. These barriers also suffer from finite-size error due to the use of a unit cell instead of a larger supercell. However, since CL-20 has four molecules in the unit cell and 36 atoms per molecule, the simulations of a larger supercell were too costly to converge at this point.

We used density functional theory (DFT) to study the relative stability of the three known polymorphs in comparison to the intermediate form that was observed in the MetaD simulation. We performed density functional theory (DFT) calculations employing the generalized gradient approximation (GGA) using the Perdew-Burke-Ernzerhof (PBE)\cite{Perdew1996} exchange-correlation functional in Quantum Espresso.\cite{Giannozzi2009}  Rappe-Rabe-Kaxiras-Joannopoulos (RRKJ)\cite{Rappe1990} plane wave ultrasoft pseudopotentials were employed. As CL-20 molecules make hydrogen bonds and Van der Waals interactions, we employed the empirical dispersion correction method DFT-D by Grimme.\cite{Grimme2010} A 2x2x2 Brillouin zone sampling Monkhorst-Pack\cite{Monkhorst1976} grid was used, with a cutoff energy of 40 Ry and a cutoff charge density of 400 Ry. Structural relaxations of the crystals were performed using the Broyden, Fletcher, Goldfarb, and Shannon (BFGS) algorithm.  
As the most stable polymorph, we set the energy of \(\epsilon\)-CL-20 to zero and calculate the energy differences in comparison to it. Structural relaxations were performed on each of the crystals. The lattice parameters and energies of the known polymorphs showed small changes relative to experimental values (Table S5), which demonstrates the reliability of our DFT calculations. The relative energies of \(\beta\)- and \(\gamma\)-CL-20  were 3.0 $kcal \, mol^{-1}$ and 2.2 $kcal \, mol^{-1}$, respectively,  in agreement with previous calculations by Kholod et al.\cite{Kholod2007} and experimental results.\cite{Nielsen1998}
The relaxation of the intermediate form I converged successfully. Form I has a relatively high density of 2.072 $gr \, cm^{-3}$ (close to \(\epsilon\)-CL-20) but with an relative energy of 12.5 $kcal \, mol^{-1}$ (Table S6). The DFT calculations confirm that we were able to discover a new (albeit high energy) metastable intermediate structure of CL-20.  

To conclude, we propose a simple procedure for obtaining the FES underlying conformational polymorphism in molecular solids. We utilized data from short, unbiased simulations of isolated conformers that correspond to each polymorph to successfully build CVs that drive polymorphic transitions in solid-state simulations using MC-HLDA. In the challenging case of CL-20, with three ambient polymorphs, we were able to observe frequent transition between all forms in MetaD trajectories. We reconstructed the FES from the biased simulations and found that the transitions between the known polymorphs occur through a previously unknown intermediate form. Importantly, we demonstrated that CVs derived from unbiased simulations of isolated molecules considering only the local molecular conformation are enough to drive solid-state transitions with the lattice parameters following the conformation change without being biased directly. The free energy differences between \(\epsilon\), \(\gamma\), and \(\beta\) polymorphs from the FES agree with previous experiments and calculations. The procedure described in this Letter can be applied to other molecular crystals exhibiting conformational polymorphism. It can help design of methods to isolate desired polymorphs experimentally.

\begin{acknowledgement}

B.H. Acknowledges support by Rafael Science Foundation (Grant No. 347960906), the USA-Israel Binational Science Foundation (Grant No. 2020083) and the Israel Science Foundation (Grants No. 1037/22 and 1312/22). O.E. thanks the Israel Academy of Sciences and Humanities Postdoctoral Fellowship Program for Israeli Researchers, the Faculty of Exact Sciences Dean’s Fellowship and the Ratner Center Fellowship at Tel Aviv University.
\end{acknowledgement}

\begin{suppinfo}

The Supporting Information is available free of charge 

\end{suppinfo}

\bibliography{refs}
\end{document}



\section{}
MD simulations were carried out using LAMMPS and PLUMED2. We used a time step of $1 \, fs$, which showed energy conversion within $0.1\%$ in an NVE simulation. In simulations of isolated molecules, the cell consisted of a single molecule of 36 atoms with the initial configuration of the CIF file after minimization (conjugate gradient). Then, we collected data on the fluctuations of the improper angles at a temperature of 50K and a pressure of $1 \, atm$. We used a Nose'-Hoover chains thermostat and barostat with Parrinello-Rahman full-cell fluctuations.\cite{Shinoda2004,Tuckerman2006,Parrinello1998,Martyna1998} The temperature and pressure were maintained with damping parameters of $10 \, fs$ and $100 \, fs$, respectively. Simulation of the solid consisted of four molecules (144 atoms) with the initial geometries as in the CIF file after minimization and thermalization of $1.1 \, ns$. Since LAMMPS flips the simulation box if any side length increases to 1.6 times its initial value, the cell angles can jump to complimentary angles. We reversed this process when flipping occurred in all Figures for clarity. In WTMetaD simulations, upper and lower harmonic walls with a spring constant of 35 $kcal \, mol^{-1}$ were employed to improper angles 2, 3, and 5. The upper and lower values that were employed are the minimum and maximum values observed from unbiased simulations of the isolated molecules.   
\begin{table} [H]
  \caption{Lattice parameters of CL-20 polymorphs modeled with the SB-CL20-CCNN force field, at 300 K and 1 atm}
  \label{tblS:notes}
 \begin{tabular}{lllllllll}
    \hline
    CL-20 &   & a({\AA}) & b({\AA}) & c({\AA}) & \(\alpha\)($^\circ$) & \(\beta\)($^\circ$) & \(\gamma\)($^\circ$) & density($g\cdot cm^{-3}$) \\
    \hline
    \(\epsilon\)   & Exp\textsuperscript{\emph{a}} & 8.852 & 12.556& 13.386& 90& 106.82& 90& 2.044   \\
    & CL-20-CCNN FF& 9.023 & 12.365 & 13.86 & 88.59 & 107.26 & 90.98 & 1.971 \\
    & dev \% & 1.93 & 1.52 & 3.6 & 1.6 & 0.4 & 1.1 & 3.5\\
    \(\beta\)  & Exp\textsuperscript{\emph{a}}  & 9.676 &	13.006 &	11.649 & 90 & 90 & 90 & 1.985\\
    & CL-20-CCNN FF& 9.673 & 12.374 & 12.443 & 90.5 & 91.52 & 89.25 &	1.955 \\
    & dev \% & 0.03 & 4.9 & 6.8 & 0.55 & 1.7 & 0.8 & 1.5\\
    \(\gamma\)   & Exp\textsuperscript{\emph{a}}& 13.231 & 8.17 &	14.876 & 90 & 109.17 & 90 & 1.916\\
    & CL-20-CCNN FF& 13.409 &	7.665 &	15.325 & 90.62 & 108.99 & 89.55 &	1.955 \\
    & dev \% & 1.34& 6.19 & 3.02 &	0.69& 0.16 & 0.50 & 2.0\\
    \hline
  \end{tabular}
  \textsuperscript{\emph{a}} experimental values obtained from ref~\cite{Nielsen1998}.
\end{table}

\begin{figure} [H]
  \includegraphics [scale=1.1] {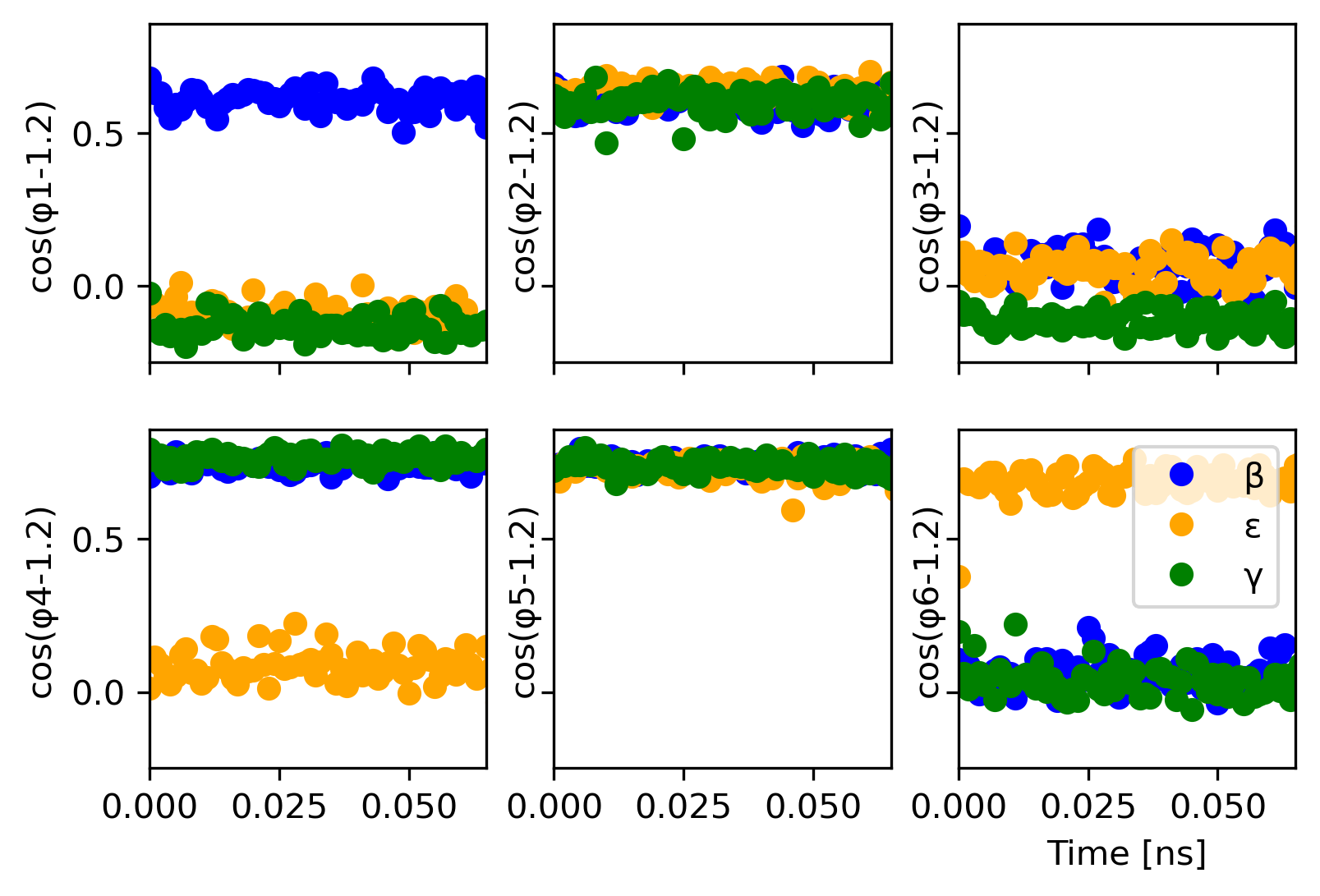} 
  \caption{Cosines of the improper angles with an offset phase of 1.2 radians from unbiased $~70 \, ps$ simulations of isolated  CL-20 conformers.}
  \label{fgr:1S}
\end{figure}

\begin{table} [H]
  \caption{Coefficients and weights of each improper angle descriptors ($\cos(\varphi_i-1.2)$) from MC-HLDA using all six of them.}
\label{tb2S:notes}
  \begin{tabular}{lllllll}
    \hline
    $\varphi_i$ & $\varphi_1$ & $\varphi_2$& $\varphi_3$& $\varphi_4$& $\varphi_5$ & $\varphi_6$ \\
    \hline
    $c_i$ & -0.4712 & -0.0417& 0.0834 & -0.8313 & -0.0899& 0.2646\\
    $w_i(\%)$ & 22.21 &	0.17& 0.7& 69.11 & 0.81& 7.0\\
    \hline
  \end{tabular}
\end{table} 

\begin{table} [H]
  \caption{Coefficients and weights ($w_i$) for improper angle descriptors ($\cos(\varphi_i-1.2)$) from MC-HLDA using only $\varphi_1$, $\varphi_4$ and $\varphi_6$.}
\label{tb3S:notes}
  \begin{tabular}{lllllll}
    \hline
    $CV_i$ & $c_1$ & $c_4$ & $c_6$& $w_1 [\%]$ & $w_4 [\%]$ & $w_6 [\%]$ \\
    \hline
    $CV_1$& -0.4342 & -0.8641 & 0.2546 & 18.8 & 74.7 & 6.5\\
    $CV_2$& -0.822 & 0.565 & -0.074& 67.6& 31.9& 0.5\\
    \hline
  \end{tabular}
\end{table}

\begin{figure} [H]
  \includegraphics [scale=1.1] {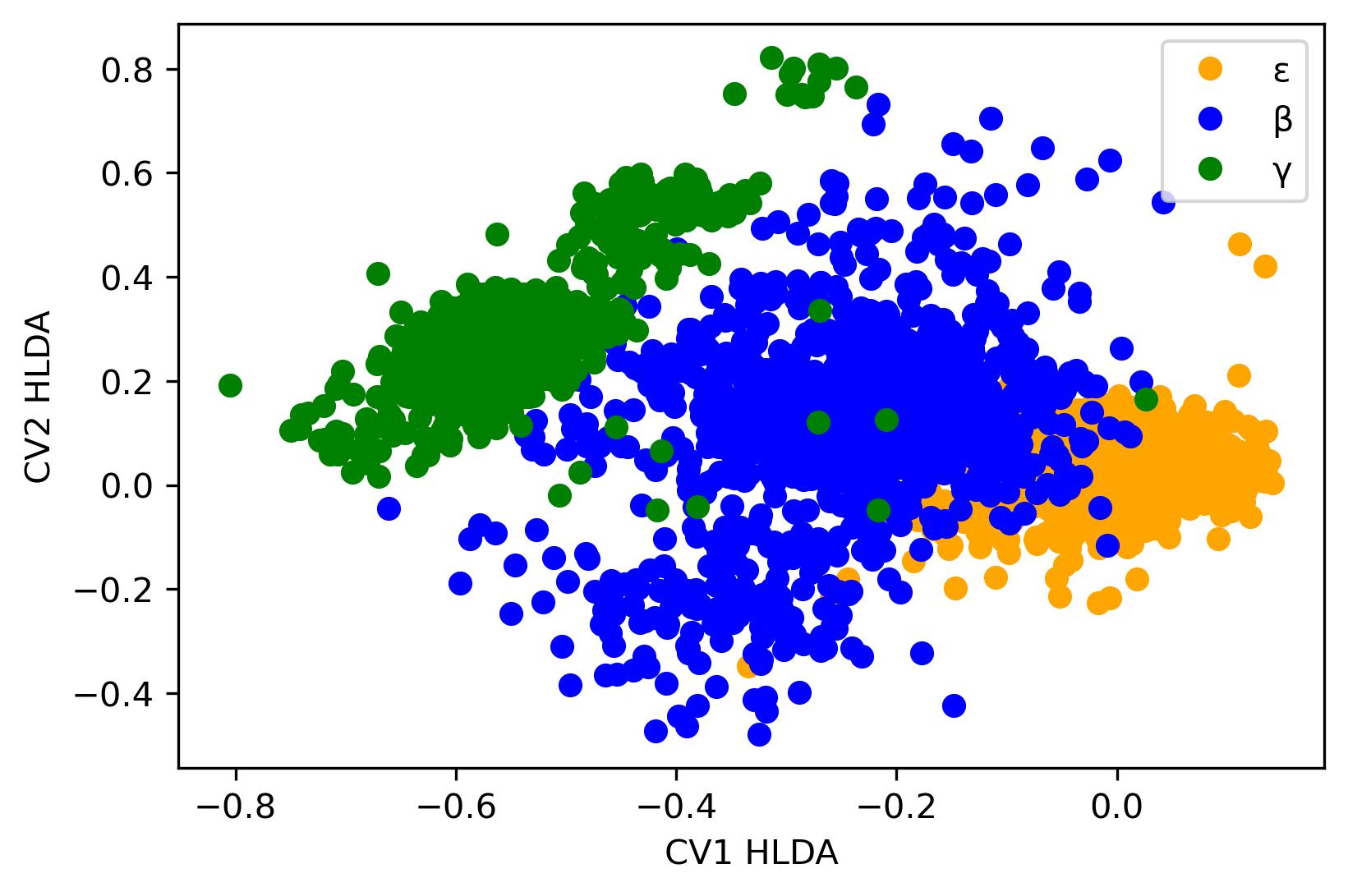} 
  \caption{Scatter plot of CL-20 polymorphs in the space of the two MC-HLDA CVs, data obtained from $\sim 70 \ ps$ unbiased unit-cell simulations.}
  \label{fgr:2S}
\end{figure}

\begin{figure} [H]
  \includegraphics [scale=0.51] {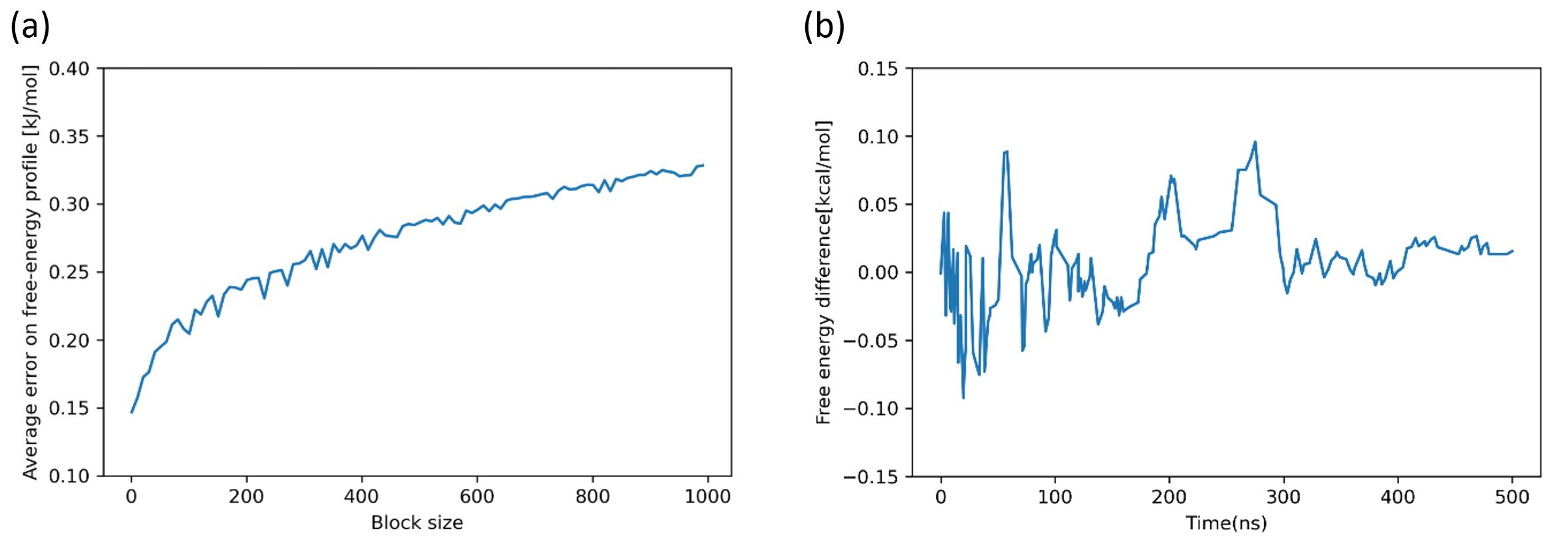} 
  \caption{(a) Average error in FES as function of block size using CV1 and, (b) energy difference between \(\epsilon\)- and \(\gamma\)-CL-20.}
  \label{fgr:3S}
\end{figure}
\begin{table} [H]
  \caption{Lattice parameters of the CL-20 defect forms}
\label{tb4S:notes}
  \begin{tabular}{llllllll}
    \hline
    Form & $a[\r{A}]$ & $b[\r{A}]$ & $c[\r{A}]$ & $a[^{\circ}]$& $b[^{\circ}]$& $c[^{\circ}]$ \\
    \hline
    II,IV & 8.909& 12.2302& 13.438& 80.421& 117.323& 94.53\\
    III & 7.997& 14.304& 16.034& 120.444& 111.43& 86.764& \\
    \(\gamma\)$_{defect}$ & 8.918& 13.083& 12.448& 90 & 90 & 90\\
    \hline
  \end{tabular}
\end{table} 

\begin{figure} [H]
  \includegraphics [scale=0.70] {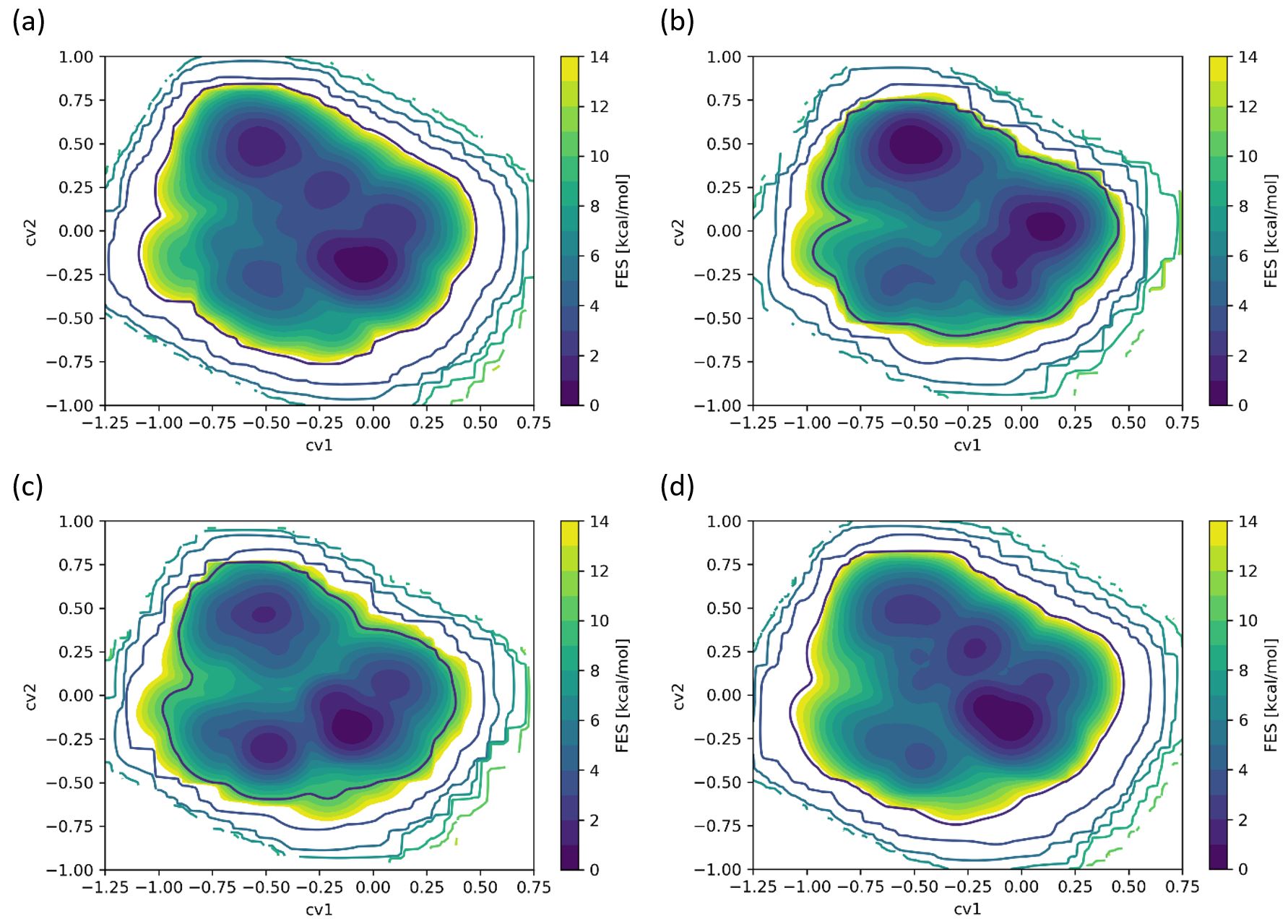} 
  \caption{FES of independent simulations with different seeds. Each seed was randomly chosen. The simulations were run for $150-500 \, ns$. }
  \label{fgr:4S}
\end{figure}

\begin{table} [H]
  \caption{Lattice parameters and densities of CL-20 polymorphs from DFT calculations}
\label{tb5S:notes}
  \begin{tabular}{llllllll}
    \hline
    $Polymorphs$ & $a[\r{A}]$ & $b[\r{A}]$ & $c[\r{A}]$ & $a[^{\circ}]$& $b[^{\circ}]$& $c[^{\circ}]$ & $\rho[gr\,cm^{-3}]$ \\
    \hline
    \(\beta\)$_{experiment}$ & 9.676& 13.006& 11.649& 90& 90& 90& 1.985\\ 
    \(\beta\)$_{DFT}$ & 9.515& 13.063& 11.419& 90& 90& 90& 2.048\\
    \(\beta\)$_{error}[\%]$ & 1.66& 0.44& 1.97& 0& 0& 0& 3.2\\
    \(\gamma\)$_{experiment}$ & 13.231&	8.170&	14.876&	90&	109.15&	90&	1.916\\ 
    \(\gamma\)$_{DFT}$ & 13.085& 8.170&	14.675&	90&	109.012& 90& 1.96\\
    \(\gamma\)$_{error}[\%]$ & 1.1& 0& 1.35& 0& 0.13& 0& 2.3\\
    \(\epsilon\)$_{experiment}$ & 8.852& 12.556& 13.386& 90& 106.82& 90& 2.044\\ 
    \(\epsilon\)$_{DFT}$ & 8.805 &12.474& 13.259& 90& 106.34& 90& 2.08\\
    \(\epsilon\)$_{error}[\%]$ & 0.53& 0.65& 0.95& 0& 0.45& 0& 1.8\\
    \hline
  \end{tabular}
\end{table} 
We applied DFT to investigate the stability of all observed intermediate and defect forms. The \(\gamma\)-form with slightly different lattice parameters observed in our MetaD simulations did not converge during cell relaxation. Thus, it is possible this form is an artifact of the applied force field.
\begin{table} [H]
  \caption{Lattice parameters and densities of the stable CL-20 intermediate form (I) from DFT calculations}
\label{tb6S:notes}
  \begin{tabular}{llllllll}
    \hline
    $Form$ & $a[\r{A}]$ & $b[\r{A}]$ & $c[\r{A}]$ & $a[^{\circ}]$& $b[^{\circ}]$& $c[^{\circ}]$ & $\rho[gr\,cm^{-3}]$ \\
    \hline
    I$_{MD-initial}$ & 9.128& 12.186& 14.578& 80.421& 117.323& 94.53& 2.015\\ 
    I$_{DFT}$ & 8.851& 12.378& 14.269& 81.336& 114.818&	94.323&	2.0728\\
    I$_{error}[\%]$ & 3.04& 1.57& 2.12& 1.14& 2.14& 0.22& 2.83\\
    \hline
  \end{tabular}
\end{table}

\bibliography{refsi}